\documentclass[10pt]{article}
\pdfoutput=1
\usepackage{jheppub,amsmath,amssymb}
\usepackage{enumerate}
\usepackage{bm}
\usepackage{bbm}
\usepackage{enumitem}
\usepackage[usenames,dvipsnames]{xcolor}
\usepackage{amsmath}
\usepackage{amsfonts}
\usepackage{amssymb}
\usepackage{graphicx}
\usepackage{hhline}
\usepackage{subfig}
\usepackage{hyperref}
\usepackage{color}
\usepackage{verbatim}
\usepackage{amsmath,amsthm,amssymb}
\usepackage{lscape}
\usepackage{float}
\usepackage{caption}
\usepackage{lscape}
\usepackage{breqn}
\usepackage{multicol}
\usepackage{graphics}
\usepackage{tikz,todonotes}
\usepackage[utf8]{inputenc}
\usepackage{autobreak}
\usepackage{verbatim}
\usetikzlibrary{arrows,positioning,decorations.markings,decorations.pathmorphing,calc}
\pgfdeclarelayer{edgelayer}
\pgfdeclarelayer{nodelayer}
\usetikzlibrary{decorations.pathreplacing}
\pgfsetlayers{edgelayer,nodelayer,main}
\tikzset{none/.style={draw=none}}
\tikzset{new edge style 2/.style={black}}
\tikzset{new style 0/.style={black}}
\tikzset{rednode/.style={draw=none, scale=0.3pt,fill=red,circle, draw}}
\tikzset{redline/.style={line width=0.3mm,red}}
\tikzset{greyE/.style={line width=0.1mm,gray}}
\usepackage[utf8]{inputenc}
\usetikzlibrary{arrows,positioning,decorations.markings,decorations.pathmorphing,calc}

\definecolor{hyperref}{RGB}{026,028,087}

\newcommand{\beq}{\begin{equation}}
\newcommand{\eeq}{\end{equation}}
\newcommand{\bea}{\begin{eqnarray}}
\newcommand{\eea}{\end{eqnarray}}
\def\be{\begin{equation}}
\def\ee{\end{equation}}

\def\beq{\begin{equation}}
\def\eeq{\end{equation}}

\newcommand{\mpl}{M_{\rm Pl}}

\newcommand{\K}{\mathcal K}

\renewcommand{\L}{\mathcal L}

\def\be{\begin{equation}}
\def\ee{\end{equation}}
\def\ba{\begin{eqnarray}}
\def\ea{\end{eqnarray}}

\def\nn{\nonumber}

\usepackage[normalem]{ulem}

\def\ba{\begin{eqnarray}}
\def\ea{\end{eqnarray}}

\def\H{\mathcal{H}}

\def\L{\mathcal{L}}
\def\K{\mathcal{K}}

\def\stu{St\"uckelberg }

\newcommand{\Ostro}{{Ostrogradski }}

\def\mn{_{\mu \nu}}
\def\mnup{^{\mu \nu}}
\def\ab{_{\alpha \beta}}
\def\abup{^{\alpha \beta}}
\def\mupn{^\mu_{\phantom{\mu}\nu}}
\def\({\left(}
\def\){\right)}

\def\mpl{M_{\rm Pl}}
\def\p{\partial}
\def\ie{{\em i.e. }}

\def\Pro{Procanuevo}
\def\Pros{Procanuevo }
\def\X{\mathcal{X}}
\def\Z{\mathcal{Z}}

\begin{document}

\title{New class of Proca interactions}

\author[a,b]{Claudia de Rham,}
\author[a]{Victor Pozsgay}
\affiliation[a]{Theoretical Physics, Blackett Laboratory, Imperial College, London, SW7 2AZ, U.K.}
\affiliation[b]{CERCA, Department of Physics, Case Western Reserve University, 10900 Euclid Ave, Cleveland, OH 44106, USA}

\emailAdd{c.de-rham@imperial.ac.uk}
\emailAdd{v.pozsgay19@imperial.ac.uk}

\abstract{We propose a new class of Proca interactions that enjoy a non--trivial constraint and hence propagates the correct number of degrees of freedom for a healthy massive spin--1 field. We show that the scattering amplitudes always differ from those of the Generalized Proca. This implies that the new class of interactions proposed here are genuinely different from the Generalized Proca and there can be no local field redefinitions between the two. In curved spacetime,  massive gravity is the natural covariantization but we show how other classes of covariantizations can be considered.}

\maketitle


\section{Introduction}

Ever since its original formulation, General Relativity (GR) has been tirelessly tested and so far, experiments and predictions agree to an unexpected precision. GR is one of the most successful physical theories but it leaves some cosmological questions unanswered. Indeed, the Universe's expansion can be explained by the introduction of dark matter in addition to a cosmological constant but its value is not technically natural. Despite decades of efforts no fully satisfying argument has been proposed to tackle the cosmological constant problem \cite{Weinberg:1988cp}. This motivates the study of modified theories of gravity as well as  theories endowed  with additional degrees of freedom. A scalar field can indeed lead to an accelerated expansion while preserving a homogeneous and isotropic matter distribution. In this context, the Galileon was introduced in \cite{Nicolis:2008in} and the Generalized Galileon in \cite{Deffayet:2009mn} as the most general interactions for a scalar field that remain free from \Ostro instabilities. It turns out that Galileons were introduced much earlier in the context of scalar--tensor theories by Horndeski \cite{Horndeski:1974wa} and are ubiquitous to many models of modified gravity at large distances \cite{Dvali:2000hr,Luty:2003vm,Nicolis:2004qq,Nicolis:2008in,deRham:2009rm,deRham:2010gu,deRham:2010eu,deRham:2010ik}.
Finally, pure Galileon interactions exhibit an interesting property at the quantum level: the interactions are stable under quantum corrections \cite{Luty:2003vm,Nicolis:2004qq,Nicolis:2008in,deRham:2010eu,Burrage:2010cu,Burrage:2011bt,deRham:2012ew,deRham:2012az}
which makes them technically natural.\\

Following this idea, modifications of General Relativity were then extended to Galileon--like theories of gauge--invariant arbitrary $p$--forms in \cite{Deffayet:2010zh, Deffayet:2016von,Deffayet:2017eqq}. Nevertheless, a no--go theorem was discovered proving that there is no non--trivial gauge--invariant Galileon--like $1$--form theory in four dimensions. Interestingly, dropping out gauge--invariance and promoting the gauge vector field to a massive Proca field, it becomes possible to construct derivative self--interactions for such a massive spin--1 field without \Ostro instabilities and thus propagating only three physical degrees of freedom. Such theories, classified under the name of  Generalized Proca (GP), or sometimes vector--Galileons, were thoroughly investigated in \cite{Heisenberg:2014rta, Tasinato:2014eka, Allys:2015sht, Allys:2016jaq, Jimenez:2016isa}. GP is the most general Lagrangian for a massive spin--1 whose equations of motion for both the helicity--1 and --0 modes remain second order in derivatives hence ensuring that the theory only propagates three degrees of freedom in four dimensions, see~\cite{Heisenberg:2018vsk} for a review. Within this framework, a proof of the uniqueness of the GP action can be derived. An interesting property of GP is the fact that the pure \stu field interactions precisely coincide with the generalized Galileon ones. \\

Since its formulation GP has had a huge impact on cosmology and gravity. Such theories have been considered for applications to both astrophysical systems \cite{Chagoya:2016aar, Minamitsuji:2016ydr, Cisterna:2016nwq,Chagoya:2017fyl,Heisenberg:2017hwb, Minamitsuji:2017aan,Heisenberg:2017xda,Kase:2017egk, deFelice:2017paw,Kase:2018owh,Rahman:2018fgy,Kase:2018voo,Kase:2020yhw} and cosmology \cite{DeFelice:2020sdq, Oliveros:2019zkl, Emami:2016ldl, Nakamura:2018oyy, Kase:2018nwt, Domenech:2018vqj, Heisenberg:2016wtr, DeFelice:2016uil, DeFelice:2016yws, Allys:2015sht}. The screening of the GP `fifth force' was considered in \cite{Heisenberg:2014rta} and  \cite{Nakamura:2017lsf}. Beyond--GP interactions were considered in~\cite{Heisenberg:2016eld,Seifert:2019xan,GallegoCadavid:2019zke} and could potentially lead to remarkable effects on deficit angles \cite{Heisenberg:2016lux} and cosmology \cite{Nakamura:2017dnf}. The related question of interactions within the context of tensor--vector theories was explored in \cite{Kimura:2016rzw}. Non--abelian interactions were considered in \cite{Allys:2016kbq,Gomez:2019tbj} with an application to cosmology in \cite{Rodriguez:2017wkg}, and the generalization to multiple spin--1 fields in \cite{Jimenez:2016upj,ErrastiDiez:2019trb, ErrastiDiez:2019ttn}. The constraint algebra was re--investigated in   \cite{Jimenez:2019hpl} and the relation to the vector fields that enter the decoupling limit of massive gravity in AdS in \cite{deRham:2018svs}. The quantum consistency of these classes of interactions within the context of quantum effective field theories were also considered in  \cite{Amado:2016ugk,deRham:2018qqo,Ruf:2018vzq}. Scalar and vector galileons are reviewed in \cite{Rodriguez:2017ckc}.\\

Upon constructing the GP set of interactions \cite{Heisenberg:2014rta, Tasinato:2014eka}, an important implicit ingredient is that the equations of motion for both the helicity--0 and --1 modes of the massive spin--1 field remain at most second order in derivatives. This assumption appears to be related to the requirement that the constraint is uniquely determined by the equation of motion with respect to the component  $A_0$ of the vector field\footnote{This specific assumption is not explicitly formulated as such in the generic formalism of \cite{ErrastiDiez:2019trb} but other implicit assumptions on how the constraint ought to manifest itself effectively reduce the formalism to the same type of GP interactions.}.  Under this assumption, the theory is indeed unique as shown. Phrased in this way however it is natural to explore whether the constraint could manifest itself differently while preserving the correct number of degrees of freedom. The analogue of this possibility was successfully explored within the context of massive gravity \cite{deRham:2010kj}, first considered in \cite{deRham:2010gu} and implemented in \cite{deRham:2011rn}. The possibility was then also later implemented within the context of scalar--tensor theories, coming under the name of `Beyond--Horndeski'
\cite{Gleyzes:2014dya,Zumalacarregui:2013pma,Langlois:2015cwa,Langlois:2015skt} and further degenerate higher--order theories (DHOSTs) were considered in \cite{Langlois:2015cwa,Achour:2016rkg,Crisostomi:2016tcp,Crisostomi:2016czh,Ezquiaga:2016nqo,Motohashi:2016ftl}. Implementations of constraints can indeed be subtle in theories with multiple fields as highlighted in \cite{deRham:2011qq,deRham:2016wji}. With this perspective in mind, in this paper, we shall consider a new type of Proca interactions which manifest a constraint and hence only propagate three dynamical degrees of freedom in four spacetime dimensions, but differ from the standard GP interactions. Since massive gravity has provided an original framework for exploring non--trivial implementations of constraints, it shall serve us as a guiding tool in constructing consistent fully non--linear Proca interactions and will allow us to prove the existence of a new type of massive spin--1 field theory  that is free of \Ostro instabilities, and propagates the required number of degrees of freedom.\\

 The rest of the manuscript is organized as follows: in Section~\ref{sec:RevGP}, we start by reviewing the GP interactions and provide their expression in the decoupling limit. We then introduce the full non--linear action for our proposed new Proca interactions or {\it \Pro}  in Section~\ref{sec:NewProca} before giving its perturbative expansion up to quartic order and its decoupling limit. We also prove that the theory carries a constraint and give a non--perturbative and explicit formula for the null eigenvector of the Hessian matrix.
 A by--product of this analysis is a generic proof of the absence of ghost of massive gravity in the \stu language. Such proof was indicated in \cite{deRham:2011rn} and carried out explicitly for a specific model in \cite{Hassan:2012qv}. On the other hand, the method provided in Section~\ref{sec:NewProca} is general and carries beyond that specific model.
 In Section~\ref{sec:Ineq}, we compute tree--level $2 \rightarrow 2$ scattering amplitudes and conclude that the S--matrix of GP never coincides with that of \Pros no matter the choice of coefficients, hence proving that both types of theories genuinely differ. We discuss the coupling of \Pros to gravity in Section~\ref{sec:CouplGrav}. We end with an outlook in Section~\ref{sec:Outlook}. Appendix~\ref{app:Null} provides the details proving the existence of a null eigenvector for any class of \Pros theory hence proving the existence of a constraint. Details used to compute the $2 \rightarrow 2$ scattering amplitude are given in Appendix~\ref{sec:appKin}. Finally Appendix~\ref{sec:appDL} provides explicit expressions for the vector--scalar interactions that arise in the decoupling limit of massive gravity.

 Throughout this paper, we work in four flat spacetime dimensions with mainly positive $(-+++)$ signature, unless specified otherwise.

\section{Review of Generalized Proca}
\label{sec:RevGP}

\subsection{Formulation}
\label{ssec:GP}

Generalized Proca is the most general theory of a massive vector field $A_{\mu}$ including an arbitrary number of derivative self--interactions such that its equations of motion remain second order and is free of \Ostro instabilities when including the helicity--0 part $\phi$ of the \stu field $A_\mu \to A_\mu +\p_\mu \phi/m$. This property  ensures that the theory has three propagating degrees of freedom in four dimensions\footnote{As we shall see the requirement that the equation of motion for $\phi$ remains second order in derivatives is a sufficient condition for the absence of \Ostro instabilities but not always a necessary one.}. In this language,  the helicity--0 mode $\phi$ is then nothing other than a Galileon. \\

Requiring the equations of motion to be at most second order in derivatives implies that GP interactions include at most one derivative per field at the level of the action and are hence solely expressed in terms of $A_\mu$ and $\p_\mu A_\nu$. In deriving the full action, it is useful to separate out the gauge--invariant building blocks \ie the Maxwell strength field $F\mn$ and its dual $\tilde F\mn$ and the gauge--breaking contributions that involve the \stu field $\phi$. One can then parameterize the GP Lagrangians in terms of the powers of the gauge--breaking contribution $\p A$. The advantage of this ordering is that it is finite in the sense that \textit{all} the interactions are listed, and the remaining infinite freedom is captured by arbitrary functions. In this language, we have \cite{Heisenberg:2014rta}
\begin{equation}
	\L_{\text{GP}} = \sum_{n=2}^6 \L_n\,,
\label{eq:LGP}
\end{equation}
where,
\begin{align}
	\L_2 &= f_2(A_{\mu}, F\mn, \tilde{F}\mn)
		\label{eq:LGP2} \\
	\L_3 &= f_3(A^2) (\partial \cdot A)
		\label{eq:LGP3} \\
	\L_4 &= f_4(A^2) [(\partial \cdot A)^2 - \partial_{\mu} A_{\nu} \partial^{\nu} A^{\mu}]
		\label{eq:LGP4} \\
	\L_5 &= f_5(A^2) [(\partial \cdot A)^3 -3 (\partial \cdot A) \partial_{\mu} A_{\nu} \partial^{\nu} A^{\mu} + 2 \partial_{\mu} A_{\nu} \partial^{\nu} A^{\rho} \partial_{\rho} A^{\mu} ] + \tilde{f}_5(A^2) \tilde{F}^{\mu \alpha} \tilde{F}^{\nu}_{\phantom{\nu} \alpha} \partial_{\mu} A_{\nu}
		\label{eq:LGP5} \\
	\L_ 6 &= \tilde{f}_6(A^2) \tilde{F}^{\mu \nu} \tilde{F}^{\alpha \beta} \partial_{\alpha} A_{\mu} \partial_{\beta} A_{\nu}\,.
		\label{eq:LGP6}
\end{align}
All the functions $f_n$'s and $\tilde{f}_n$'s are arbitrary polynomial functions so these Lagrangians span an infinite family of operators depending on the form of these functions\footnote{Notice that this formulation differs ever so slightly with that originally introduced in \cite{Heisenberg:2014rta}. For instance the contribution to $\L_4$ proportional to $c_2$ in Eq.~(2.2) of \cite{Heisenberg:2014rta} is here absorbed into the function $f_2$, however both formulations are entirely equivalent.}. For comparison with other theories, and to compute scattering amplitudes, it is convenient to expand all the functions $f_n$ and $\tilde{f}_n$ in the most generic possible way and repackage the Lagrangian \eqref{eq:LGP} perturbatively in a field expansion. In this case, the theory is expressed perturbatively as
\begin{equation}
	\L_{\text{GP}} = \sum_{n=2}^\infty \frac{1}{\Lambda_2^{2(n-2)}}\L_{\text{GP}}^{(n)} \,,
\label{eq:LGPbis}
\end{equation}
where $\Lambda_2$ is introduced as the dimensionful scale  for the interactions and
where up to quartic order
\begin{align}
	\L_{\text{GP}}^{(2)} &= -\frac{1}{4}F\mnup F\mn - \frac{1}{2}m^2 A^2
		\label{eq:LGP2bis} \\
	\L_{\text{GP}}^{(3)} &=  a_1 m^2 A^2 \partial_{\mu} A^{\mu} + a_2 \tilde{F}^{\mu \alpha} \tilde{F}^{\nu}_{\phantom{\nu} \alpha} \partial_{\mu} A_{\nu}
		\label{eq:LGP3bis} \\
	\L_{\text{GP}}^{(4)} &=  b_1 m^4 A^4 + b_2 m^2 A^2 F\mnup F\mn + b_3 m^2 A^2 \left[ (\partial \cdot A)^2 - \partial_{\alpha} A_{\beta} \partial^{\beta} A^{\alpha} \right] + b_4 m^2 F^{\mu \alpha}F^{\nu}_{\phantom{\nu} \alpha} A_{\mu} A_{\nu}  \nonumber \\
													 & + b_5 F\mnup F\abup F_{\mu \alpha} F_{\nu \beta} + b_6 F\mnup F\mn F\abup F\ab + b_7 \tilde{F}\abup \tilde{F}\mnup \p_{\alpha} A_{\mu} \p_{\beta} A_{\nu}\,,
		\label{eq:LGP4bis}
\end{align}
with the coefficients $a_{i}$ and $b_j$ being dimensionless constants. The scaling is introduced so as to `penalize' the breaking of gauge--invariance with the scale $m$ (see \cite{deRham:2018qqo} for the appropriate scaling of operators in gauge--breaking effective field theories).
Note that there exists various different but equivalent ways to express the Lagrangian perturbatively depending on how total derivatives are included, nevertheless irrespectively on the precise formulation,  there exists $2$ linearly independent terms at cubic order and $7$ at quartic order (ignoring total derivatives).

\subsection{Generalized Proca in the Decoupling Limit}
\label{ssec:GPDL}

For any theory, its decoupling limit (DL) is determined by scaling parameters of the theory so as to be able to focus on the irrelevant operators that arise at the lowest possible energy scale while maintaining all the degrees of freedom alive in that limit. Hence by definition, the number of degrees of freedom remains the same in the DL. Taking a DL is different from taking a low--energy effective field theory and also differs from switching off interactions or degrees of freedom. See for instance Refs.~\cite{deRham:2014wfa,deRham:2014zqa,deRham:2016wji} for more details on the meaning of a DL. \\

In the particular case of GP, the DL is taken by first introducing the \stu field explicitly in a canonically normalized way,
\begin{equation}
	A_{\mu} \rightarrow A_{\mu} + \frac{1}{m}\partial_{\mu} \phi\,,
\label{eq:Stuck}
\end{equation}
so that the kinetic term for the helicity--0 mode is explicitly manifest in \eqref{eq:LGP2bis}, indeed $\L^{(2)}_{\rm GP}\supset -\frac 12 (\p \phi)^2$. We then take the DL by sending the mass $m$ to zero and $\Lambda_2\to \infty$ in such a way as to keep the lowest interaction scale finite in that limit. Denoting generic interactions scales $\Lambda_p$ by $\Lambda_p=(m^{p-2}\Lambda_2^2)^{1/p}$  (with $\Lambda_3\equiv (m \Lambda_2^2)^{1/3}$), one can check that the lowest scale at which interactions appear is $\Lambda_3$. The $\Lambda_3$-DL of GP is then taken by sending
\begin{equation}
	m \rightarrow 0, \quad \Lambda_2 \rightarrow \infty \quad \text{keeping} \quad \Lambda_3 \equiv  (m \Lambda_2^2)^{1/3} = \text{const.}\,,
\label{eq:Lambda3DL}
\end{equation}
once all the fields are properly normalized.

Upon taking this DL, one notices that out of all the interactions that entered the quartic GP Lagrangian $\L_{\text{GP}}^{(4)} $ in  \eqref{eq:LGP4bis} only terms proportional to $b_3$ and $b_7$ survive and one ends up with
\begin{equation}
	\L_{\text{DL GP}} = \L_{\text{DL GP}}^{(2)} + \frac{1}{\Lambda_3^3} \L_{\text{DL GP}}^{(3)} + \frac{1}{\Lambda_3^6} \L_{\text{DL GP}}^{(4)} + \frac{1}{\Lambda_3^9} \L_{\text{DL GP}}^{(5)}\,,
\label{eq:LDLGP}
\end{equation}
where the first four Lagrangians are given by
\ba
	\L_{\text{DL GP}}^{(2)} &=& -\frac{1}{4}F\mnup F\mn - \frac{1}{2} (\partial \phi)^2
		\label{eq:LDLGP2} \\
	\L_{\text{DL GP}}^{(3)} &=& a_1 (\p \phi)^2 [\Phi] + a_2 \tilde{F}^{\mu \alpha} \tilde{F}^{\nu}_{\phantom{\nu} \alpha} \Phi\mn
		\label{eq:LDLGP3}
= a_1 (\p \phi)^2 [\Phi] + a_2 F^{\mu \alpha} F^{\nu}_{\phantom{\nu} \alpha} \(\Phi\mn - \frac{1}{2} [\Phi]\eta\mn \)
		\label{eq:LDLGP3b} \\
	\L_{\text{DL GP}}^{(4)} &=& b_3 (\p \phi)^2 \([\Phi]^2-[\Phi^2]\) + b_7 \tilde{F}^{\alpha \beta} \tilde{F}\mnup \Phi_{\alpha\mu}\Phi_{\beta\nu}
		\label{eq:LDLGP4} \\
&=& b_3 (\p \phi)^2 \([\Phi]^2-[\Phi^2]\)
+ b_7 F^{\alpha \beta} F\mnup \Bigg[ \Phi_{\mu\alpha}\Phi_{\nu \beta}
+2 \eta\ab \left( \Phi^2\mn - \Phi\mn [\Phi] \right)
+\frac{1}{2} \eta_{\mu \alpha}\eta_{\nu \beta} \left( [\Phi]^2 - [\Phi^2] \right) \label{eq:LDLGP4b} \Bigg]\,,\nn\hspace{-0.5cm} 										
\ea
where we used the notation $\Phi\mn=\p_\mu \p_\nu \phi$.
In contrast with the $9$ parameters family of interactions up to quartic order for GP, its DL up to quartic order only includes the cubic and quartic Galileon interactions as well as two genuine mixings between the helicity--0 and --1 modes, parameterized by $a_2$ and $b_7$. The quintic Lagrangian $\L_{\text{DL GP}}^{(5)}$ involves the quintic Galileon and can include interactions between the helicity--0 and --1 modes although the precise form of these interactions is not relevant for this study.

\section{\Pro}
\label{sec:NewProca}

\subsection{Full non--linear theory}
\label{sssec:NonLinTh}

We shall now build our intuition from massive gravity to derive a new type of fully non--linear Proca interactions. The DL of massive gravity includes an infinite number of scalar--vector interactions whose exact form was provided in \cite{Ondo:2013wka}. Interestingly, the scalar--vector sector of the DL of massive gravity can in principle be thought of as the DL of a Proca theory, similarly to what was considered in Sec.~\ref{ssec:GPDL} for GP. On another hand, the scalar--vector interactions included in the DL of massive gravity involve higher derivatives acting on the fields and thus violate the original assumption in deriving the most general GP operators. Yet massive gravity has been proven to be ghost--free in many different languages \cite{deRham:2010kj,Hassan:2011hr,deRham:2011rn,deRham:2011qq,Hassan:2012qv} and hence so is its DL. Indeed, as emphasized in \cite{deRham:2011rn,deRham:2011qq,deRham:2016wji} the constraint can manifest slightly differently in theories with multiple fields and the existence of higher derivatives in the equations of motion does not necessarily imply an \Ostro ghost instability. For instance, there can be a linear combination of the equations of motion which is free from higher derivatives so that no higher--order \Ostro ghost instability occurs \cite{deRham:2011qq,deRham:2016wji}. This phenomenon is similar to what is observed in Beyond--Horndeski theories and other extensions \cite{Gleyzes:2014dya,Zumalacarregui:2013pma,
Langlois:2015cwa,Langlois:2015skt,Achour:2016rkg,Crisostomi:2016tcp,
Crisostomi:2016czh,Ezquiaga:2016nqo,Motohashi:2016ftl,deRham:2016wji}. \\

Massive gravity is the theory of an interacting massive spin-2 field $h\mn$. In terms of a gravitational dynamical  metric $g\mn$, the spin-2 field $h\mn$ is expressed as $\mpl^{-1}h\mn=g\mn-\eta\mn$ in unitary gauge. The fact that the Minkowski metric $\eta\mn$ is not diffeomorphism invariant implies that expressed in this way $h\mn$ is not a tensor. However gauge invariance can be easily restored through the introduction of four \stu fields $\phi^a$ which transform as scalars under coordinate transformations. Indeed, expressed in terms of the tensor $f\mn$
\ba
\mpl^{-1}h\mn&=&g\mn-f\mn\\
{\rm with}\quad f\mn&=&\eta_{ab}\p_\mu \phi^a \p_\nu \phi^b\,,
\label{eq:fmn1}
\ea
the quantity $h\mn$ is now a tensor under diffeomorphisms.  In the limit where $\mpl \to \infty$ we may identify the index $a$ as a Lorentz index.  Splitting the fields $\phi^a$ as $=x^a+A^a$, the field $A^a$ can then be associated with a Lorentz vector which is anchored in the very formulation of massive gravity. \\

However, at this stage, the link between massive gravity and Proca interactions is not necessarily immediately manifest as massive gravity always includes the tensor modes.
 In fact there is no limit of {\it pure} massive gravity that would lead to a massive vector theory on Minkowski. Indeed, for such a  limit to occur, the helicity-0 mode of the massive spin-2 field of massive gravity should play the role of the helicity-0 mode of the massive vector field. However in {\it pure} massive gravity on Minkowski, the helicity-0 mode only acquired its kinetic term from mixing with the tensor mode \cite{deRham:2010ik}. \\

 Instead one can consider the DL of massive gravity on AdS \cite{deRham:2018svs} where the helicity--0 mode acquires its own kinetic term without the need for a coupling with the tensor modes. Alternatively one can consider generalized massive gravity \cite{deRham:2014lqa,deRham:2014gla} where the scalar mode also acquires its own kinetic term. In both cases a new type of $\Lambda_2$--decoupling limit that only involves couplings between the scalar and vector modes can be considered \cite{deRham:2015ijs,deRham:2016plk,Gabadadze:2017jom,deRham:2018svs,Gabadadze:2019lld}. In \cite{deRham:2018svs} it was shown that on AdS, the resulting scalar--vector interactions could never be expressed as a local and Lorentz invariant field redefinition of the scalar--vector interactions that arise in the DL of GP, suggesting that these classes of interactions were indeed distinct from GP. In what follows we shall build from these results to provide a new class of non--linear ``\Pro" massive Proca interactions that rely on the same structure as the decoupling limit of massive gravity.
%
%
We start with a Lorentz vector field $A_\mu$ and just as was the case in the GP theory of section~\ref{sec:RevGP} we continue working on flat spacetime with the Minkowski metric $\eta\mn$ (coupling to gravity is considered in section~\ref{sec:CouplGrav}).\\

 These considerations are mainly motivational for this context and following our intuition from massive gravity, we may consider the tensor $f\mn$ defined in \eqref{eq:fmn1} where the $\phi^a$'s are expressed in terms of the vector field as follows
 \ba
\label{eq:phia}
\phi^a=x^a+\frac{1}{\Lambda_2^2}A^a\,,
\ea
so that in terms of the vector field, the quantity $f\mn$ is expressed as \footnote{The object $f\mn$ is simply a Lorentz tensor constructed out of the first derivative of the Lorentz vector $A_\mu$ and at this level has no connection with any type of auxiliary metric. Note that in this context of a massive vector field, introducing the quantity $\phi^a$ in terms of the coordinate $x^a$ may be misleading as it suggests a breaking of Poincar\'e invariance, however, the quantity we shall be interested in, $f\mn$, is manifestly a Poincar\'e tensor if $A_\mu$ is itself a Poincar\'e vector as is clear from the expression \eqref{eq:deff2}.}
\ba
	f\mn[A] = \eta\mn + 2 \frac{\p_{(\mu} A_{\nu)}}{\Lambda_2^2} + \frac{\p_{\mu} A_\alpha \p_{\nu} A_\beta \eta^{\alpha \beta}}{\Lambda_2^4}\,.
		\label{eq:deff2}
\ea
Next we introduce the (Poincar\'e) tensor  $\K\mupn$ defined as
\ba
\label{eq:defK1}
 \K\mupn &=& \X\mupn -\delta \mupn\\
 \text{with }	\quad \X\mupn[A] &=& \left( \sqrt{\eta^{-1}f[A]}  \right)\mupn \qquad \text{i.e.}\qquad \X^{\mu}_{\phantom{\mu} \alpha} \X^{\alpha}_{\phantom{\alpha} \nu} = f\mupn = \eta^{\mu \alpha}f_{\alpha \nu}\,,
		\label{eq:defK2}
\ea
where, in the gravitational context,  $\K\mupn$ would be playing the role of the extrinsic curvature \cite{deRham:2013awa,deRham:2014zqa} and $\X$ that of the vielbein \cite{Hinterbichler:2012cn}.

In four dimensions, the theory of the vector field $A_\mu$ we propose is then expressed as
\begin{equation}
	\L_{\K}[A] = \Lambda_2^4 \sum_{n=0}^4 \alpha_n(A^2) \L_n[\K[A]]\,,
\label{eq:defLK}
\end{equation}
where the order by order Lagrangians are defined as usual by
\begin{equation}
	\L_n[\K] = \epsilon^{\mu_1 \cdots \mu_n \mu_{n+1} \cdots \mu_4} \epsilon_{\nu_1 \cdots \nu_n \mu_{n+1} \cdots \mu_4} \K^{\nu_1}_{\phantom{\nu_1}\mu_1} \cdots \K^{\nu_n}_{\phantom{\nu_n}\mu_n}\,.
\label{eq:defLnK}
\end{equation}
More explicitly, we have
\begin{align}
	\L_0[\K] &= 4! \label{eq:defL0K} \\
	\L_1[\K] &= 3! [\K] \label{eq:defL1K} \\
	\L_2[\K] &= 2!([\K]^2 - [\K^2]) \label{eq:defL2K} \\
	\L_3[\K] &= [\K]^3 - 3[\K][\K^2] + 2[\K^3] \label{eq:defL3K} \\
	\L_4[\K] &= [\K]^4 - 6[\K]^2[\K^2] + 3[\K^2]^2 + 8[\K][\K^3] - 6[\K^4] \label{eq:defL4K}\,,
\end{align}
and we use the standard notation for the trace, $[\K] = \text{tr}(\K)$. As mentioned before, the theory \eqref{eq:defLK} has no gravitational degrees of freedom, rather it is a pure vector theory with an infinite tower of self--interactions. We shall prove in section~\ref{ssec:Hessian} that this vector--field theory corresponds to a Proca theory with at most three propagating degrees of freedom.

 Note that $\L_0$ is just a potential for the vector field, $\alpha_0(A^2)\L_0=V(A^2)$, which is where the vector field will carry its mass from and so it is essential for the consistency of this theory that $\alpha_0$ includes at the very least a contribution going as $\alpha_0\supseteq -\frac 12 (m^2/\Lambda_2^4) A^2$.

\subsection{Perturbative Action}
\label{sssec:PertAction}

The exact non--perturbative Lagrangian is expressed in \eqref{eq:defLnK} but it is instructive to consider its  perturbative expression and we shall provide it up to quartic order in the field (as needed for the $2 \rightarrow 2$ tree--level scattering amplitudes).
To provide such a perturbative expression, we first Taylor expand the functions $\alpha_n(A^2)$ as follows
\begin{equation}
	\alpha_n(A^2) = \bar{\alpha}_n + \frac{m^2}{\Lambda_2^4} \bar{\gamma}_n A^2 + \frac{m^4}{\Lambda_2^8} \bar{\lambda}_n A^4 + \cdots\,.
\label{eq:betans}
\end{equation}
Plugging it into \eqref{eq:defLK} and requiring the canonical normalization for the quadratic Lagrangian (Maxwell with a mass term) requires the following normalization:
\ba
\bar \alpha_1=-\frac 13 \( 1 - 2 \bar \alpha_2 \) \qquad {\rm and}\qquad \bar \gamma_0=-\frac 1{48}\,.
\ea
The perturbative expansion up to quadratic order then takes the form
\begin{equation}
	\L_{\K} = \L_{\K}^{(2)} + \frac{1}{\Lambda_2^2} \L_{\K}^{(3)} + \frac{1}{\Lambda_2^4} \L_{\K}^{(4)} + \cdots\,,
\label{eq:LK}
\end{equation}
with
\ba
	\L_{\K}^{(2)} &=& -\frac{1}{4}F\mnup F\mn - \frac{1}{2}m^2 A^2
		\label{eq:LK2} \\
	\L_{\K}^{(3)} &=& \frac 14 \(2\bar \alpha_2-3 \bar \alpha_3\) [F^2] [\p A]
+ \frac14\(1-4 \bar \alpha_2+6\bar \alpha_3\) F^2\mn \p^{\mu} A^{\nu} +6 \bar \gamma_1 m^2 A^2 [\p A]\label{eq:LK3} \\
	\L_{\K}^{(4)} &=& \frac 1{32}\(\bar \alpha_2-3 \bar \alpha_3+6 \bar \alpha_4\) [F^2]^2
+ \frac{1}{64}\(5-20 \bar \alpha_2-12 \bar \alpha_3+168 \bar \alpha_4\) F^2\mn F^2{}^{\mu\nu} \label{eq:LK4} \\
&+& \frac 38 \(\bar \alpha_3-4 \bar \alpha_4\) [F^2] \left( [\p A]^2 - \p_{\alpha}A_{\beta} \p^{\beta} A^{\alpha} \right)
-\frac 18 F^2\mn \p^{\beta} A^{\mu} \p_{\beta} A^{\nu} \nonumber \\
&+&\(\frac 12 \bar \alpha_2+\frac 34 \bar \alpha_3-6 \bar \alpha_4\)F^2{}^{\mu\nu}\(\p^{\beta} A_{\mu} \p_{\beta} A_{\nu}-[\p A]\p_\mu A_\nu\)\nn\\
&+&\(-\frac 18+\frac 12 \bar \alpha_2-3 \bar \alpha_4\)F^{\mu\nu}F^{\alpha \beta}\p_\mu A_\alpha \p_\nu A_\beta\nn\\
&+&m^2 A^2 \left[2 \bar \gamma_2 [\p A]^2-\(\frac 32 \bar \gamma_1+\bar \gamma_2\)\p_\mu A_\nu \p^\nu A^\mu
+\(\frac 32 \bar \gamma_1-\bar \gamma_2\)\p_\mu A_\nu \p^\mu A^\nu\right]\nn\\
&+&24 \bar \lambda_0 m^4 A^4\nn\,,
\ea
where we use the notation $F^2\mn=F_{\mu}{}^{\alpha}F_{\nu \alpha}$ and $[F^2]=F^{\mu\nu}F\mn$\,.

\subsection{Decoupling Limit}
\label{sssec:NewDL}
It will also be instructive to consider the DL of this \Pros theory. Introducing the helicity--0 \stu field $\phi$ as in \eqref{eq:Stuck}
using the same scaling as in \eqref{eq:Lambda3DL}, we get
\begin{equation}
	\L_{\K {\text{DL}}} = \L_{\K {\text{DL}}}^{(2)} + \frac{1}{\Lambda_3^3} \L_{\K {\text{DL}}}^{(3)} + \frac{1}{\Lambda_3^6} \L_{\K {\text{DL}}}^{(4)} + \cdots\,,
\label{eq:LKDL}
\end{equation}
with
\ba
	\L_{\K {\text{DL}}}^{(2)} &=& -\frac{1}{4}F\mnup F\mn - \frac{1}{2} (\p \phi)^2
		\label{eq:LKDL2} \\
	\L_{\K {\text{DL}}}^{(3)} &=& \frac 14 \(2\bar \alpha_2-3 \bar \alpha_3\) [F^2] \Box \phi
+ \frac14\(1-4 \bar \alpha_2+6\bar \alpha_3\) F^2\mn \Phi^{\mu\nu} +6 \bar \gamma_1 (\p \phi)^2 \Box \phi
		\label{eq:LKDL3} \\
	\L_{\K {\text{DL}}}^{(4)} &=&
\frac 38 \(\bar \alpha_3-4 \bar \alpha_4\) [F^2] \left( [\Phi]^2-[\Phi^2] \right)
-\frac 18 F^2\mn \Phi^2{}^{\mu\nu} \label{eq:LKDL4}
 \\&+&
\(\frac 12 \bar \alpha_2+\frac 34 \bar \alpha_3-6 \bar \alpha_4\)F^2{}^{\mu\nu}\(\Phi^2\mn-[\Phi]\Phi\mn\)\nn\\
&+&\(-\frac 18+\frac 12 \bar \alpha_2-3 \bar \alpha_4\)F^{\mu\nu}F^{\alpha \beta}\Phi_{\mu \alpha}\Phi_{\nu \beta}\nn\\
&+&2 \bar \gamma_2 (\p\phi)^2\([\Phi]^2-[\Phi^2]\)\nn\,.
\ea
In this DL, we see that the coefficients $\bar \gamma_{1,2}$ govern the pure cubic and quartic Galileon interactions while the other $\bar \alpha_{2,3,4}$ coefficients govern the interactions between the vector and the scalar sector. This scalar--vector mixing matches precisely those that arise in the DL of massive gravity \cite{Ondo:2013wka} up to a trivial redefinition of the coefficients (see Appendix \ref{sec:appDL}). While the DL of GP truncates at quintic order (see Eq.~\eqref{eq:LDLGP}), we note that the DL of \Pros does not truncate and involves an infinite number of interactions in the scalar--vector sector. Moreover one can check that these interactions are never exactly of the GP form even after local and Lorentz invariant field redefinitions \cite{deRham:2018svs}.\\

While GP was constructed so as to ensure that its DL leads to second order equations of motion one can check explicitly that the \Pro's DL involves higher derivatives in its equations of motion. At first sight, one may worry that those higher derivatives are related to \Ostro ghost--like instabilities however we shall see below that the constraint remains in the \Pros theory and in four dimensions, only three degrees of freedom are excited. Since the theory enjoys the same vacuum as a free Proca theory with no ghost, this ensures that there can be no ghost excitations when working about configurations that are connected to the standard Proca vacuum when remaining within the regime of validity of the theory. In what follows we start by proving that the Hessian in two dimensions has a vanishing eigenvalue. We then prove the existence of a null eigenvector for the Hessian in arbitrary dimensions, hence signaling the existence of a constraint. We note that since we are dealing with a parity preserving Lorentz--invariance theory, there can be no half number of propagating degrees of freedom and hence the existence of a primary second class constraint automatically ensures the existence of a secondary constraint (see Ref.~\cite{deRham:2014zqa} for more details on that point).

\subsection{Hessian}
\label{ssec:Hessian}

We shall now show that the Hessian of \Pros always includes a vanishing eigenvalue hence implying the existence of a constraint that removes the would--be \Ostro ghost.

\subsubsection{Example}
\label{sssec:2d}

To start with, we may consider the theory in two dimensions and focus on the Lagrangian given by
\ba
	\L^{(\text{2d})} = -2[\K] -\frac 12 m^2 A^2\,.\label{eq:L2d}
\ea
 In two dimensions, an interactive massive vector field could in principle excite two degrees of freedom, but a healthy Proca theory should only excite one. We shall thus determine the Hessian of \Pros in two dimensions and prove that it only involves one non--vanishing eigenvalue. For simplicity we define
\ba
	x =\frac{1}{\Lambda_2^2} \p_\mu A^\mu \qquad {\rm and }\qquad
    y =\frac{1}{\Lambda_2^2} F_{01}=\frac{1}{\sqrt{2}\Lambda_2^2} \sqrt{-[F^2]}\,. \label{eq:xdef}\label{eq:ydef}
\ea
Then the Lagrangian takes the very simple form
\begin{align}
	\L^{(\text{2d})} = -2[\K] -\frac 12 m^2 A^2&= - 4 \sqrt{1 + x + \frac{x^2-y^2}{4}} +4 -\frac 12 m^2 A^2 \\
														&= - 4 \sqrt{1 + \frac{\p_{\mu} A^{\mu}}{\Lambda_2^2} + \frac{2(\p_{\mu} A^{\mu})^2 + [F^2]}{8\Lambda_2^4}}+4 -\frac 12 m^2 A^2\,,\label{eq:L2db}
\end{align}
and the Hessian matrix is given by
\ba
 \mathcal{H}^{ab} =  \frac{\p^2 \L}{\p \dot A_a \p \dot A_b}
= \frac{2}{[\X]^3 \Lambda_2^4}
		\begin{pmatrix}
			y^2 &  y (2+x) \\
			 y(2+x) & (2+x)^2
		\end{pmatrix}\,.
\label{eq:Hessmat}
\ea

It is a straightforward to check that the determinant of the Hessian does indeed vanish, signaling that one of the vector components  is non--dynamical and leaving only one propagating degree of freedom in two dimensions. The null eigenvector simply reads
\begin{equation}
v_a =
		\begin{pmatrix}
			1 \\
			0
		\end{pmatrix}+\frac{1}{2}
		\begin{pmatrix}
			x \\
			-y
		\end{pmatrix}\,.
\label{eq:eigvec}
\end{equation}
We see that this null eigenvector is perturbatively connected with the vector $(1,0)$ and still ensures that $A_0$ is not dynamical. Next we shall prove the existence of a similar type of null eigenvector for any \Pros theory in any number of dimensions.

\subsubsection{Null Eigenvector in arbitrary dimensions}
\label{sssec:Null4d}

We shall now give a non--perturbative proof of the absence of ghost in four or any other dimensions, for the full theory, by deriving analytically the Hessian matrix and giving an expression for a null eigenvector. The proof for the absence of ghost follows from the arguments provided in \cite{deRham:2011rn,deRham:2014lqa,deRham:2014gla,deRham:2016plk} and generalizes the proof given in \cite{Hassan:2012qv} beyond the minimal model. We recall that $\K = \X-1$ with $\X=\sqrt{\eta^{-1}f}$ and we introduce the matrix $\Z$ defined as
\begin{equation}
\Z = \X^{-1} \eta^{-1}\,.
\label{eq:defZ}
\end{equation}
One can check that $\Z$ is symmetric, using the same similarity transformation as introduced in \cite{deRham:2014naa},
\ba
\Z^{-1}=\eta \X= \(\eta \sqrt{\eta^{-1}f}\eta^{-1}\)\eta
=\sqrt{f \eta^{-1}}\eta=\X^T \eta=\(\Z^{-1}\)^{T}\,.
\ea
It follows that $\Z=\Z^T$ and
\begin{align}
	\Z\abup f_{\beta \gamma} &= \X^{\alpha}_{\phantom{\alpha} \gamma} \label{eq:Zf} \\
	\Z\mnup f_{\nu \alpha} \Z\abup &= \eta^{\mu \beta} \label{eq:ZfZ}
\end{align}
Now if we evaluate the $00$-component of \eqref{eq:ZfZ} and differentiate it with respect to the time--derivative of the vector field $\dot A^a$, we find
\begin{equation}
\frac{\p }{\p \dot A^a }\(\Z^{0 \mu} f_{\mu \nu} \Z^{\nu 0} \)= 2\frac{\p \Z^{0 \mu}}{\p \dot A^a }f_{\mu \nu} \Z^{\nu 0}+\frac{2}{\Lambda_2^{2}} \Z^{00}\Z^{0\mu}\p_\mu \phi_a=0
\quad \Rightarrow \quad
\frac{\p \Z^{0 \mu}}{\p \dot A^a } \X^0_{\phantom{0} \mu} = - \Lambda_2^{-2} \Z^{00} V_a\,,
\label{eq:dZ}
\end{equation}
where $\phi_a=\eta_{ab}\phi^b$ is introduced in \eqref{eq:phia} and where we have introduce the normalized time--like vector $V_a$  defined as
\begin{equation}
	V_a = \Z^{0 \mu} \p_{\mu} \phi_a\,,
\label{eq:Va}
\end{equation}
so that $V^a V_a = -1$.
It is then straightforward to show that
\begin{equation}
	\p_{\mu} \phi_a V^a = \X^0_{\phantom{0} \mu}\,.
\label{eq:dphiV}
\end{equation}
Using these relations, we find the following expressions for the generic derivatives,
\ba
\label{eq:dK1}
	\frac{\partial}{\partial \dot{A}^a} [\X^n] = n \Lambda_2^{-2} \(\X^{n-2}\)^0{}_\mu \, \partial^{\mu} \phi_a\,,
\ea
for any $n\ge 1$. In particular for $n=1$, this implies $\frac{\partial}{\partial \dot{A}^a} [\X] = \Lambda_2^{-2}V_a$.
Now that every element has been introduced, we can compute the momenta first and then the Hessian matrices for each order in $\K$ or $\X$. Since $\K$ and $\X$ are linearly related to one another, the $\L_n[\K]$ can be expressed as linear combinations of the $\L_n[\X]$ as summarized in \cite{deRham:2014zqa} and we may use either choice for the following argument without loss of generality. We will then show that $V_a$ is actually the null eigenvector for the Hessian derived for any linear combination of  $\L_n[\K]$ or equivalently any linear combination of $\L_n[\X]$ hence proving the existence of a constraint. \\

Let us start with the easiest case by considering $\L_1[\X]$. The conjugate momentum associated to $\phi_a$ is already given in \eqref{eq:dK1} and we have
\begin{equation}
	p_a^{(1)} =\Lambda_2^{4}\frac{\p \L_1[\X]}{\p \dot A^a}=\Lambda_2^{2} V_a\,.
\label{eq:pa1}
\end{equation}
The Hessian associated with this Lagrangian is then
\ba
\mathcal{H}_{ab}^{(1)}=\Lambda_2^{4}\frac{\p^2 \L_1[\X]}{\p \dot A^a \p \dot A^b}=\Lambda_2^{2}\frac{\p V_a}{\p \dot A^b}\,.
\ea
Rather than computing this Hessian explicitly, it is actually easier to simply make use of the property of $V_a$ (and the fact that it has constant norm),
\ba
\Lambda_2^{-2}	\mathcal{H}_{ab}^{(1)}V^a = \frac{\partial V_a}{\partial \dot{A}^b}V^a
	                          = \frac{1}{2} \frac{\partial (V_a V^a)}{\partial \dot{A}^b}
											      = \frac{1}{2} \frac{\partial (-1)}{\partial \dot{A}^b}
											      = 0\,,
\label{eq:eig1}
\ea
hence proving that $V_a$ is indeed a null eigenvector of $\H^{(1)}_{ab}$.\\

Generalizing this result for any \Pros Lagrangian is straightforward and the details are provided in appendix~\ref{app:Null}, where we show that for any Lagrangian of the form \eqref{eq:defLK}, the associated Hessian carries the {\it same} null eigenvalue $V_a$
for all linear combinations of Lagrangians  $\L_n[\X]$.
It follows that any linear combination of $\L_n[\X]$ or $\L_n[\K]$ carries a constraint and only excites three degrees of freedom in four dimensions. Interestingly, the way the constraint manifests itself differs from the way it does in GP (their respective null eigenvectors differ). This implies that considering a hybrid theory composed of GP {\it and} \Pros interactions would not enjoy a constraint.\\

Remarkably, the existence of a constraint is now manifest irrespectively of the choices of $\alpha_n$. The argument provided here, therefore, extends prior proofs for the absence of ghost in massive gravity in the \stu language beyond what was proposed in \cite{deRham:2011rn} and \cite{Hassan:2012qv}. Such a general proof was previously missing in the literature. Interestingly with the exact form of the null eigenvector at hand, one should now be able to determine the full non--linear version of the \stu field in terms of which massive gravity and \Pros can be manifestly expressed in first order form.

\section{Inequivalence with Generalized Proca}
\label{sec:Ineq}

The aim of this Section is to show that the \Pros theory provided in \eqref{eq:defLK} does not enter the scope of GP. It is clear that \Pros includes an infinite number of operators with arbitrarily high order in $(\p A)$ while GP only includes a finite number of those (putting aside the gauge--invariant interactions). However by itself, this does not imply that both theories may not still be the same in disguise for instance through a sophisticated field redefinition or even an analogue to the Galileon duality proposed in \cite{deRham:2013hsa,deRham:2014lqa}. In \cite{deRham:2018svs} it was shown that on AdS, there were no local and Poincar\'e invariant field redefinitions between GP and the DL of massive gravity. In what follows we shall show that this result is generic, and even account for more subtle types of space--dependent field redefinitions like generalized Galileon dualities, there can be no local field redefinition that maps GP with \Pros theories. This will be done in full generality by computing and comparing the S matrix of both theories in section~\ref{ssec:scattering} but to start with we shall start by recalling that the very way the constraint gets satisfied differs in GP and \Pros theories as can be seen very easily in two dimensions.

\subsection{Appetizer}
\label{ssec:naive2d}

By definition, a GP is a theory carrying a constraint and thus propagating only $d-1$ degrees of freedom in $d$ spacetime dimensions. However the existence of a constraint can take various different forms and the non--dynamical variable does not necessarily need to be  $A_0$ itself, it may be a linear combination of $A_0$ and other components of the vector field. In GP, the Hessian is always of the form
\begin{equation}
 \tilde{\mathcal{H}}^{\rm (GP)}_{ab} =
		\begin{pmatrix}
			0 & 0 \\
			0 & \#
		\end{pmatrix}\,.
\label{eq:Hessmat2}
\end{equation}
In \Pro, on the other hand, while the Hessian still carries a null eigenvalue, its form differs from \eqref{eq:Hessmat2} at least when expressed in terms of the components of the field $A_\mu$,
indeed, the Hessian for the two--dimensional Lagrangian $\L^{\rm(2d)}$ \eqref{eq:L2d} is expressed in \eqref{eq:Hessmat} and is not of the form \eqref{eq:Hessmat2} even though both Hessians have null determinant. \\

 Let us now suppose there could exist a field redefinition $A_\mu \to \tilde{A}_\mu(A)$ such that the Hessian for $\tilde A$ is of the form \eqref{eq:Hessmat2}.
After the field--redefinition, the Hessian matrix takes the form
\begin{equation}
	\tilde{\mathcal{H}}_{ab} =  \frac{\delta \tilde{A}_a}{\delta A_c} \mathcal{H}_{cd}  \frac{\delta \tilde{A}_b}{\delta A_d}\,.
\label{eq:Hessmat3}
\end{equation}
Asking for $\tilde{\mathcal{H}}$ to be of the form \eqref{eq:Hessmat2} would require the field redefinition to be such that
\begin{equation}
	(\dot{A}_1-A_0') \frac{\delta \tilde{A}_0}{\delta A_0} = (2\Lambda_2^2+A_1'-\dot{A}_0) \frac{\delta \tilde{A}_0}{\delta A_1}\,.
\label{eq:nonlocal}
\end{equation}
which cannot be satisfied without imposing a non--local expression for $\tilde A_0$ in terms of $A_0$ and $A_1$. At this stage, one can already expect there to be no local field redefinition that brings \Pros back to a GP form. The same conclusion was highlighted in AdS in Ref.~\cite{deRham:2018svs}. We shall make this statement more rigorous in what follows.

\subsection{Scattering amplitudes}
\label{ssec:scattering}
To consolidate the previous argument on the absence of local field redefinition that would bring \Pros into a GP form, we shall compare here the tree--level $2 \rightarrow 2$ scattering amplitudes for both theories. \\

First we emphasize that at the linear level, GP and \Pros are identical, indeed $\L_{\text{GP}}^{(2)}$  in \eqref{eq:LGP2bis} is identical to $\L_{\K}^{(2)}$ in \eqref{eq:LK2}. This implies that the free asymptotic states defined in both theories are the same and one can meaningfully compare the amplitudes computed for each model.
Computing the indefinite $2 \rightarrow 2$ tree--level amplitudes in both theories is straightforward but for conciseness, we only present here the results for scatterings of some specific definite helicity states. As we shall see, these definite amplitudes are by themselves sufficient to show that the new Proca interactions we introduced in section~\ref{sec:NewProca} differ from those of GP theories.\\

For simplicity, we choose to describe the kinematic space with the Mandelstam variable $s$ (center of mass energy$^2$) and the scattering angle $\theta$, see Appendix~\ref{sec:appKin}.

Starting with $++\rightarrow--$, the respective scattering amplitudes in \Pros and GP  are given by
\begin{align}
\label{eq:ampKppmm}
	\mathcal{A}^{++\rightarrow--}_{\K}(s,\theta)
		&= -\frac{i}{64\Lambda_2^4} \left(\frac{s^3}{m^2} (1 + 4\bar{\alpha}_2 - 6 \bar{\alpha}_2)^2 \right. \\
		& \quad \quad \quad \quad - 2 s^2 \left(4(\bar{\alpha}_2 - 3 \bar{\alpha}_3 + 6\bar{\alpha}_4) + (1+ 8\bar{\alpha}_2 - 12\bar{\alpha}_3)^2 - 96(1 + 4\bar{\alpha}_2 - 6\bar{\alpha}_3) \bar{\gamma}_1 \right) \nonumber \\
		& \quad \quad \quad \quad + 8 m^2 s \left(1 + 4(\bar{\alpha}_2 - 3\bar{\alpha}_3 + 6\bar{\alpha}_4 - 12\bar{\gamma}_1 + 8\bar{\gamma}_2) + 2(4\bar{\alpha}_2 - 6\bar{\alpha}_3 - 24\bar{\gamma}_1)^2 \right) \nonumber \\
		& \quad \quad \quad \quad - 16 m^4 \left(1 - 48\bar{\gamma}_1 + 32\bar{\gamma}_2 + 768\bar{\lambda}_0 \right) \nonumber \\
		& \quad \quad \quad \quad + (s-4m^2)(-8(1 - 4\bar{\alpha}_2 + 6\bar{\alpha}_3)m^2 + 3(1 - 4\bar{\alpha}_2 + 4\bar{\alpha}_3 + 8\bar{\alpha}_4)) \sin(\theta)^2 \nonumber \\
		& \quad \quad \quad \quad - \left. \frac{4(s-m^2)^2 (s-2m^2) (s-4m^2) (1 - 4\bar{\alpha}_2 + 6\bar{\alpha}_3)^2}{4m^2(s-3m^2) + (s-4m^2)^2 \sin(\theta)^2} \sin(\theta)^2 \right)\,,\nonumber
\end{align}
and
\begin{align}
\label{eq:ampGPppmm}
	\mathcal{A}^{++\rightarrow--}_{\text{GP}}(s,\theta)
		&= -\frac{i}{4\Lambda_2^4} \left(\frac{s^3}{m^2} a_2^2 \right.  \\
		& \quad \quad \quad \quad + 8 s^2 \left(a_2 (a_1 - a_2) - b_5 - 4b_6 \right) \nonumber \\
		& \quad \quad \quad \quad + 8 m^2 s \left(-4b_2 + b_4 + 4b_5 + 16b_6 + 2(a_1-a_2)^2 \right) \nonumber \\
		& \quad \quad \quad \quad - 32 m^4 \left(b_1 - 2b_2 + 2b_5 + 4b_6 \right) \nonumber \\
		& \quad \quad \quad \quad - 4 (s-4m^2)(b_5 s + b_4 m^2) \sin(\theta)^2 \nonumber \\
		& \quad \quad \quad \quad - \left. \frac{4(s-m^2)^2 (s-2m^2) (s-4m^2) a_2^2}{4m^2(s-3m^2) + (s-4m^2)^2 \sin(\theta)^2} \sin(\theta)^2 \right)\,.\nonumber
\end{align}
Remarkably we see that perturbative unitarity gets broken when $s^3\sim \Lambda_2^4 m^2 \sim \Lambda_3^6$, hence confirming the existence of non-trivial operators at the scale $\Lambda_3$.
If both theories were equivalent they would predict the same scattering amplitudes for any incoming and outgoing polarization states. We will note any amplitude difference for a given set of polarizations $\Delta \mathcal{A}$ and ask them to vanish for all $(s,\theta)$, in particular
\begin{align}
	\Delta \mathcal{A}^{++\rightarrow--}(s,\theta) &= \mathcal{A}^{++\rightarrow--}_{\K}(s,\theta) - \mathcal{A}^{++\rightarrow--}_{\text{GP}}(s,\theta) \nonumber \\
		&= \sum_{n=0}^3 C_n s^n m^{4-2n} + (s-4m^2) \sin(\theta)^2 (C_4 m^2 + C_5 s) \label{eq:deltaA} \\
		& \quad \quad + C_6 \frac{(s-m^2)^2 (s-2m^2) (s-4m^2)}{4m^2(s-3m^2) + (s-4m^2)^2 \sin(\theta)^2} \sin(\theta)^2 \,, \nonumber
\end{align}
where the constants $C_{n}$ are expressed in terms of the coupling constants of the GP and \Pros only.
For the scatterings \eqref{eq:ampKppmm} and \eqref{eq:ampGPppmm} to be equivalent, one should have $C_n=0$ for all $n = 0, \dots, 6$.
Imposing these relations in terms of the coupling constants then sets
\begin{equation}
	\begin{cases}
		a_2 &= \pm \frac{1}{4} \\
		b_1 &=\frac 18 ( 1 - 2a_1 + 8a_1^2- 12 \bar{\gamma}_1 -288 \bar \gamma_1^2+192 \bar{\lambda}_0)\\
		b_2 &= \frac14\(2a_1^2 -3 \bar{\gamma}_1- 72\bar{\gamma}_1^2-2 \bar{\gamma}_2 \)\\
		b_4 &= \frac{1}{8} \\
		b_5 &= - \frac{1}{64} (3 - 4\bar{\alpha}_2 + 24 \bar{\alpha}_4 ) \\
		b_6 &= \frac{1}{32} (2a_1 - \bar{\alpha}_2 + 6 \bar{\alpha}_4 - 12\bar{\gamma}_1) \\
		\bar{\alpha}_3 &= \frac{2}{3} \bar{\alpha}_2
	\end{cases}
\label{eq:Cnsol}
\end{equation}
From these relations, it is clear that the most generic \Pros theory cannot be put in the form of GP since one already needs to impose $\bar{\alpha}_3 = \frac{2}{3} \bar{\alpha}_2$ but looking at other polarizations makes it clear that even within this choice of coefficients the theories are never equivalent. Indeed, turning now to $+- \rightarrow +-$ scatterings then upon imposing the solution \eqref{eq:Cnsol}, we find
\begin{equation}
	\Delta \mathcal{A}^{+-\rightarrow+-}(s,\theta=0) = \frac{i}{4\Lambda_2^4}(4m^2-s)s\,,
\label{eq:deltaAbis}
\end{equation}
at this stage there are no further couplings one can dial to ensure the equivalence and so irrespectively of the choice of coefficients  $\left\{\bar{\beta}_i,\bar{\gamma}_i,\bar{\lambda}_i,a_i,b_i\right\}$ the \emph{full} tree--level $2 \rightarrow 2$ scattering amplitude of our new Proca interactions never matches that predicted by GP. This concludes the proof that both theories are fundamentally different and are \emph{not} equivalent.

\section{(Re)coupling to gravity}
\label{sec:CouplGrav}

The covariantization of \Pros is very similar to that of the Galileon \cite{Nicolis:2008in}. Originally derived from the DL of the gravitational Dvali--Gabadadze--Porrati model \cite{Dvali:2000hr,Luty:2003vm}, the natural covariantization of the Galileon is hence the DGP model itself, or generalized massive gravity. Remarkably, it was indeed shown in Ref.~\cite{Garcia-Saenz:2019yok} that massive gravity is the natural way the Galileon symmetry can be gauged.

However taken as a scalar field in its own right, one may envisage a covariantization of the Galileon where the fields transform as a diffeomorphism (diff) scalar in the embedding gravitational theory. Such types of covariantizations lead to the `Covariant Galileon',  \cite{Deffayet:2009wt}, proxy theories of massive gravity \cite{deRham:2011by} or more generically to Horndeski \cite{Horndeski:1974wa} and where then further extended to Beyond--Horndeski and more generic classes of degenerate higher order theories \cite{Gleyzes:2014dya,Zumalacarregui:2013pma,
Langlois:2015cwa,Langlois:2015skt,Achour:2016rkg,Crisostomi:2016tcp,
Crisostomi:2016czh,Ezquiaga:2016nqo,Motohashi:2016ftl}.

Viewed as Effective Field Theories, the Galileon just like GP or \Pros have a very low cutoff at the scale $\Lambda_3$ (or lower \cite{deRham:2017xox}) and there can be a continuum of interactions between the scale $\Lambda_3$ and the Planck scale so that the question of what the natural covariantization of these theories is may not be particularly meaningful. However for many of these classes of theories, one may postulate the existence of a Vainshtein--type of mechanism that may allow us to push their regime of applicability beyond the scale $\Lambda_3$.

\subsection{Generalized Massive Gravity as the Natural Covariantization}
\label{ssec:MGCov}

As introduced in Section~\ref{sec:NewProca}, \Pros is heavily inspired by massive gravity. When considering the coupling of \Pros to gravity (or when considering \Pros in curved spacetime) a natural covariantization is therefore simply the theory of massive gravity introduced in \cite{deRham:2010kj} (or rather its generalized form introduced in \cite{deRham:2014lqa,deRham:2014gla}) where the Lorentz vector $A_\mu$ is not promoted to a diff vector (ie to a vector under general coordinate transformations) but rather is considered as being part of a diff scalar $\phi^a$ as introduced in \eqref{eq:phia}.\\

In this covariantization of  \Pro, the quantity $f\mn$ remains identical as that defined in  \eqref{eq:deff2}, still expressed in terms of the Minkowski metric,
\ba
\label{eq:fmn2}
f\mn =\p_\mu \phi^a \p_\nu \phi^b \eta_{ab}  =\eta\mn + 2 \frac{\p_{(\mu} A_{\nu)}}{\Lambda_2^2} + \frac{\p_{\mu} A_\alpha \p_{\nu} A_\beta \eta^{\alpha \beta}}{\Lambda_2^4}\,,
\ea
 even though the field is living on an arbitrary spacetime with dynamical metric $g\mn$. The metric $g\mn$ enters the definition of $\K$ which is now defined as \cite{deRham:2010kj}
\ba
\K\mupn=\(\sqrt{g^{-1}f}\)\mupn-\delta\mupn\,,
\ea
leading to the lagrangian for massive gravity including the dynamics of the metric,
\ba
\label{eq:cov}
\L_{\rm Cov}=\frac{\mpl^2}{2}\sqrt{-g}R[g]+\Lambda_2^4 \sqrt{-g}\sum_{n=0}^4 \alpha_n(\phi) \L_n[\K]\,.
\ea
This generalized theory of massive gravity reduces to \Pros in the limit where gravity is `switched off' or decoupled, $\mpl\to \infty$ so long as $\alpha_0$ includes a quadratic term in the vector field. The absence of ghost in this covariantization follows from the absence of ghost in massive gravity \cite{deRham:2010kj,Hassan:2011hr,deRham:2014lqa,deRham:2014gla}.

\begin{figure}[h]
  \centering
  \includegraphics[width=0.8\textwidth]{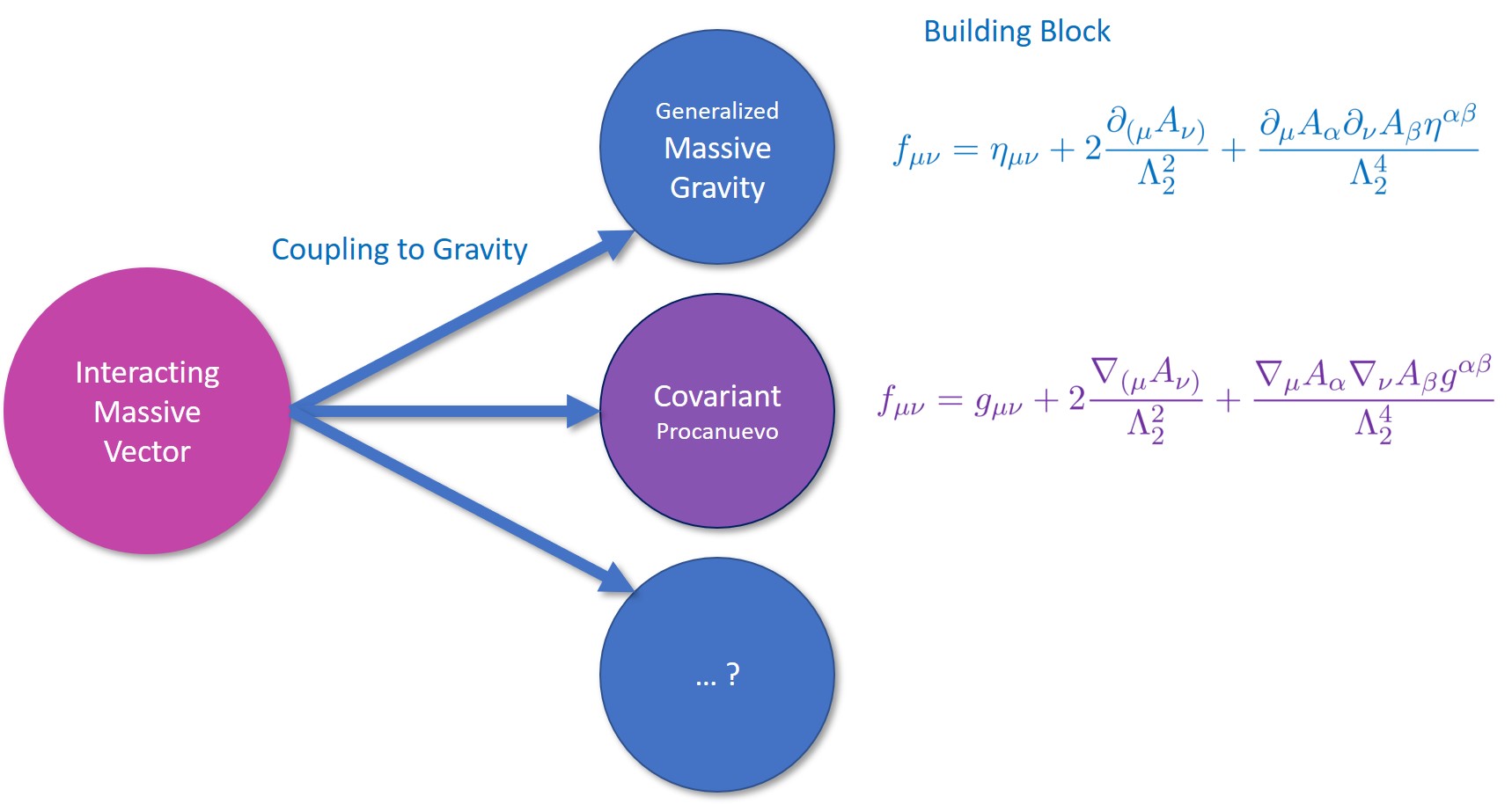}
  \caption{Any theory proposed in flat space can admit various potential classes on different covariantization. Generalized massive gravity is a natural one to consider for \Pros since this is where it was originally inspired from, but other non--equivalent covariantizations can be considered. See Ref.~\cite{deRham:2019wjj} for related arguments. }\label{Fig:Covs}
\end{figure}

\subsection{Alternative Covariantization}
\label{ssec:AltCov}

When coupling to gravity, an alternative approach is to treat $A_\mu$ as a diff vector. In doing so, instead of using the quantity $f\mn$ defined in \eqref{eq:fmn2}, the building block of the covariant theory would then be the diff tensor $f^{(g)}\mn$ defined as
\ba
\label{eq:Fmn}
f^{(g)}\mn = g\mn + 2 \frac{\nabla_{(\mu} A_{\nu)}}{\Lambda_2^2} + \frac{\nabla_{\mu} A_\alpha \nabla_{\nu} A_\beta g^{\alpha \beta}}{\Lambda_2^4}\,.
\ea
In this covariantization, the gravitational--vector theory would be given by an expression similar to \eqref{eq:cov} but with $\K$ now being a diff tensor defined as
\ba
\label{eq:KF}
\K\mupn=\(\sqrt{g^{-1}f^{(g)}}\)\mupn-\delta\mupn\,.
\ea
The absence of ghost in this covariantization is non--trivial and indeed non--minimal couplings to gravity, for instance of the form $G^{\mu\nu}A_\mu A_\nu$ may in principle need to be included to ensure the absence of \Ostro ghost. Proving the existence of such a class of covariantization which is entirely free of the \Ostro ghost is beyond the scope of this work however it can easily be done in two dimensions.\\

For concreteness, consider the covariant version of the two--dimensional Lagrangian $\L^{(\text{2d})}$ introduced in \eqref{eq:L2d},
\ba
	\L^{(\text{2d})}_{\rm cov} = \sqrt{-g}\(-2[\K] -\frac 12 m^2 A_\mu A_\nu g^{\mu \nu}\)\,,\label{eq:L2dc}
\ea
with $\K$ now as defined in \eqref{eq:KF}. This theory includes five variables that may be split into the lapse $N$, shift $n_1$ and 1--dimensional spatial metric $\gamma_{11}=\gamma$, and the two components of the vector field $A_0$ and $A_1$. For the theory to avoid any type of \Ostro ghost, out of these five variables, only one of them ought to be dynamical (in practise the helicity--0 mode of the massive vector).
To check that the theory \eqref{eq:L2dc} does indeed satisfy this property, we may compute the five--dimensional field space Hessian given by
\ba
\label{eq:Hab}
\H_{AB}=\frac{\p^2 \L^{(\text{2d})}_{\rm cov}}{\p \dot \Psi ^A \dot \Psi^B}\,,
\ea
with $\Psi^A=\{N, n_1, \gamma, A_0, A_1\}$ and check that it is of rank--1.

Upon defining the following two quantities,
\begin{align}
	B &= A_0 - A_1 n_1 \label{eq:functionB} \\
	C &= 4 N^3 \gamma^2 + 4 A_0' n_1 N \gamma^2 + 2 A_1' N \gamma (N^2 - n_1^2 \gamma) + 2 n_1' N \gamma^2 (A_0 - 2 A_1 n_1)  \\
		& \quad + (2 N' \gamma - N \gamma') (A_1(N^2 + n_1^2 \gamma) - A_0 n_1 \gamma) - 2 \dot{A}_0 N \gamma^2 + 2 A_1 \dot{n}_1 N \gamma^2 + (A_0 - A_1 n_1) (2 \dot{N} \gamma - N \dot{\gamma}) \gamma\,,\nn
\end{align}
one can check explicitly that the Hessian defined in \eqref{eq:Hab} can actually  be written in the form
\begin{equation}
	\H_{ab} =-\frac{1}{2 \gamma^5 N^8} P_a P_b\,,
\label{eq:hess2dgrav}
\end{equation}
with the field space vector $P$ defined as
\ba
P_a=\(-2 B \gamma^2 F_{01}, -2 N \gamma^2 A_1 F_{01}, N \gamma B F_{01}, C, 2 N \gamma^2 F_{01}\)\,.
\ea
This directly implies that the Hessian is of rank--1 and hence the theory \eqref{eq:L2dc} only propagates one degree of freedom in two dimensions.
This shows that the direct covariantization of the quantity $f\mn$ as in \eqref{eq:Fmn} is a `consistent choice'  in two dimensions in the sense that it maintains all the constraints required both for gravity and for the Proca field. Extending the covariantization more generically to four dimensions is beyond the scope of this work as the argument provided was merely to illustrate the presence of different types of alternative covariantizations as illustrated in Fig.~\ref{Fig:Covs}.

\section{Outlook}
\label{sec:Outlook}

In this paper, we proposed a new interactive theory for a single massive vector field with derivative self--interactions and free of \Ostro ghost instability. The \Pros Lagrangian is heavily inspired by massive gravity and is genuinely different from the GP classes of interactions. We started by proving that \Pros exhibits a constraint in two dimensions before providing the exact non--perturbative form of the null eigenvector of the Hessian matrix in any dimensions. \Pros provides an insightful example of an \Ostro ghost--free theory with a non--trivial null eigenvector. Indeed, whereas GP imposes $A_0$ to be non--dynamical, \Pro's constraint arises as a combination of $A_0$ and the spatial field components. This is already a strong hint indicating that both theories are fundamentally different. To complete the proof more rigorously, we computed the $2 \rightarrow 2$ scattering amplitudes in GP and \Pros theories and showed that they could never be matched irrespectively of the choice of coefficients. This proves that their respective $S$--matrices are different and thus \Pros cannot be related to GP by any local field redefinition.\\

Throughout this work, we have focused our analysis on the existence of a constraint and on the counting of the number of propagating degrees of freedom. In itself this question is distinct from whether or not the theory provided here can ever enjoy a standard analytic, unitary, local, Lorentz--invariant and causal high energy completion\footnote{We emphasis that the absence of such high energy completion does not necessarily rule out the existence of other consistent completions, see Refs.~\cite{Keltner:2015xda,deRham:2017xox} for relevant discussions.} although some connections were previously established for massive spin--2 interactions \cite{deRham:2018qqo} using the so--called beyond--forward positivity bounds \cite{deRham:2017zjm}. Applying the forward bounds to a specific class of spin--1 effective field theory was considered in \cite{Bonifacio:2016wcb} and implications to GP and other types of massive spin--1 effective field theories in and beyond the forward limit was considered in \cite{deRham:2018qqo}. Interestingly the positivity bounds on GP requires the introduction of very specific operators and it would be interesting to understand whether the same type of arguments applies to the theory at hand. \\

It is beyond the scope of this paper to apply this theory to cosmology but based on the impact that the scalar mode of  GP has already had on cosmology and astrophysics, we hypothesize that the helicity--0 mode of \Pros could play a similar type of role while providing different classes of signatures. It could be interesting to see if the cosmological predictions of \Pros differ significantly from the well--studied one from GP. Another obvious follow up to this paper would be the study of the general covariantization of \Pros in four (or arbitrary) dimensions and generalize the prescription to multiple interacting fields\footnote{We point however that including various species of interacting fields typically reduces the possibility for the effective field theory to enjoy a standard high energy completion \cite{Alberte:2019xfh,Alberte:2019zhd}. However, it would be interesting to diagnose whether the same type of ghost as that diagnosed in \cite{Hinterbichler:2012cn,deRham:2015cha,Alberte:2019lnd} re--appears.}. Lastly, this paper provides a new theory exhibiting a constraint non--perturbatively and could motivate a more generic study of the way constraints are satisfied in various field theories. A natural question is whether GP and \Pros are the only two types of interactions  for a  massive spin--1  that exhibit a constraint or whether other families of interactions exist \cite{WorkWithSebastian}.\\

Finally, we note that in proving the existence of a constraint for \Pro, we have generalized the proof for the absence of ghost in massive gravity in the \stu language beyond what had previously been proposed in the literature. Remarkably, we now have the full non--linear expression for the null eigenvector of the Hessian. With this knowledge at hand, one should now be able to determine the full non--linear expressions for the \stu fields in terms of which massive gravity can be express in a manifestly first order form. This is left for further studies.

\bigskip
\noindent{\textbf{Acknowledgments:}}
We would like to thank Sebastian Garcia-Saenz, Lavinia Heisenberg and Andrew Tolley for useful discussions and comments, as well as Benjamin Strittmatter  for interesting discussions during preliminary parts of this work on AdS.
 The work of CdR is supported by an STFC grant ST/P000762/1, a European Union's Horizon 2020 Research Council grant 724659 MassiveCosmo ERC--2016--COG and by a Simons Foundation award ID 555326 under the Simons Foundation's Origins of the Universe initiative, `\textit{Cosmology Beyond Einstein's Theory}'. CdR thanks the Royal Society for support at ICL through a Wolfson Research Merit Award. VP  is funded by the Imperial College President's Fellowship.

\appendix

\section{Null Eigenvector for generic \Pros theories}
\label{app:Null}

In subsection \ref{sssec:Null4d} we proved explicitly that the vector $V_a$ defined in \eqref{eq:Va} as $V_a = \Z^{0 \mu} \p_{\mu} \phi_a$ with $\Z^{-1}=\eta \X$ and $\X=\sqrt{\eta^{-1}f}$ is a null eigenvector of the Hessian associated with the Lagrangian $\L_1[\X]$. We now proceed to prove this result for every other $\L_n[\X]$.

We will not go through the derivation of this result for each order in $\X$ or $\K$ but we provide here intermediate results, \ie the momenta and Hessian matrices. \\

At any order in the Lagrangian expansion \eqref{eq:defLK}, we define
\ba
	p_a^{(n)} = \frac{\p (\Lambda_2^4 \L_n[\X])}{\p \dot{A}^a} \nn \quad {\rm and }\quad
	\mathcal{H}_{ab}^{(n)}= \frac{\p^2 (\Lambda_2^4 \L_n[\X])}{\p \dot{A}^a \p \dot{A}^b} = \frac{\p p_a^{(n)}}{\p \dot{A}^b} \,. \nn
\ea
Since the Lagrangians $\L_n[\X]$ and $\L_n[\K]$ are related by linear relations,
\ba
	\L_n[\K] = \sum_k c_{n,k} \L_k[\X] \quad \Rightarrow \quad
	\mathcal{H[\K]}_{ab}^{(n)} &= \sum_k c_{n,k} \mathcal{H[\X]}_{ab}^{(k)}\,. \nn
\ea
$\bullet$ For the Lagrangian $\L_2[\X]$, we have an associated contribution to the conjugate momentum given by
\begin{equation}
	\Lambda_2^{-2}p_a^{(2)} = 4([\X]V_a + \dot{\phi}_a)
\label{eq:pa2}
\end{equation}
resulting in a contribution to the Hessian given by
\begin{equation}
	\mathcal{H}_{ab}^{(2)} = 4\left(\Lambda_2^2[\X]\frac{\partial V_a}{\partial \dot{A}^b} + V_b V_a + \eta_{ab}\right)\,.
\label{eq:H2}
\end{equation}
Given the Hessian \eqref{eq:H2}, it is straightforward to see that $V^a$ is indeed a null eigenvector, meaning that $\mathcal{H}_{ab}^{(2)} V^a=0$
\ba
	\mathcal{H}_{ab}^{(2)} V^a = 4\left(\Lambda_2^2[\X]\frac{1}{2}\frac{\partial V_a V^a}{\partial \dot{A}^b} + V_b (V_a V^a) + V_b\right)
	                           = 0\,.
\label{eq:HV2}
\ea
$\bullet$ For the Lagrangian $\L_3[\X]$, we have an associated contribution to the conjugate momentum given by
\begin{equation}
\Lambda_2^{-2}	p_a^{(3)} = 3\left([\X]^2-[\X^2]\right)V_a + 6[\X]\dot{\phi}_a + 6 X^{0 \mu}\partial_{\mu} \phi_a
\label{eq:pa3}
\end{equation}
leading to a Hessian
\begin{equation}
	\mathcal{H}_{ab}^{(3)} = 6\left([\X](V_a V_b + \eta_{ab}) + V_a\dot{\phi}_b + V_b\dot{\phi}_a + \X^{00}\eta_{ab}\right) + 3\Lambda_2^2\left([\X]^2-[\X^2]\right)\frac{\partial V_a}{\partial \dot{A}^b} + 6\Lambda_2^2\frac{\partial X^{0 \mu}}{\partial \dot{A}^b} \partial_{\mu} \phi_a
\label{eq:H3}
\end{equation}
for which we can again explicitly check that $V^a$ is a null vector,
\begin{align}
	\mathcal{H}_{ab}^{(3)}V^a &= 6\left([\X](- V_b + V_b) -\dot{\phi}_b + V_b\dot{\phi}_a V^a + \X^{00} V_b\right) + \frac{3}{2}\Lambda_2^2\left([\X]^2-[\X^2]\right)\frac{\partial (-1)}{\partial \dot{A}^b} + 6\Lambda_2^2\frac{\partial \X^{0 \mu}}{\partial \dot{A}^b} \X^0_{\phantom{0}\mu} \nn \\
	&= 6(\X^{00} + \dot{\phi}_a V^a)V_b + 3\left(\Lambda_2^2\frac{\p f^{00}}{\partial \dot{A}^b}-2\dot{\phi}_b\right) \nn \\
	&= 0
\label{eq:VH3}
\end{align}
$\bullet$ Finally, for the Lagrangian $\L_4[\X]$, the associated  conjugate momentum is given by
\begin{equation}
\Lambda_2^{-2}	p_a^{(4)} = 4\left([\X]^3 - 3[\X][\X^2] + 2[\X^3]\right)V_a + 12\left([\X]^2-[\X^2]\right)\dot{\phi}_a + 24\left([\X]\X^{0 \mu}-f^{0 \mu}\right)\partial_{\mu}\phi_a\,,
\label{eq:pa4}
\end{equation}
leading to the Hessian
\begin{align}
	\mathcal{H}_{ab}^{(4)} =& 12\left([\X]^2-[\X^2]\right)(V_b V_a + \eta_{ab}) + 24\left([\X]\dot{\phi}_b + \X^{0\mu}\partial_{\mu}\phi_b\right)V_a + 4\Lambda_2^2\left([\X]^3 - 3[\X][\X^2] + 2[\X^3]\right) \frac{\partial V_a}{\partial \dot{A}^b} \nonumber \\
	                        &+ 24\left([\X]V_b + \dot{\phi}_b\right)\dot{\phi}_a + 24\left(\X^{0 \mu} V_b + [\X] \Lambda_2^2 \frac{\partial \X^{0 \mu}}{\partial \dot{A}^b} + \partial^{\mu}\phi_b + \eta^{0\mu}\dot{\phi}_b\right)\partial_{\mu}\phi_a \nonumber \\
&+ 24\left([\X]\X^{00}-f^{00}\right)\eta_{ab}
\label{eq:H4}
\end{align}
for which $V^a$ is yet again a null eigenvector,
\begin{align}
	\mathcal{H}_{ab}^{(4)}V^a =& 12\left([\X]^2-[\X^2]\right)(- V_b + V_b) - 24\left([\X]\dot{\phi}_b + \X^{0\mu}\partial_{\mu}\phi_b\right) + 2\Lambda_2^2\left([\X]^3 - 3[\X][\X^2] + 2[\X^3]\right) \frac{\partial (-1)}{\partial \dot{A}^b} \nonumber \\
	                           &+ 24\left([\X]V_b + \dot{\phi}_b\right)\X^{0}_{\phantom{0}0} + 24\left(\X^{0 \mu} V_b + [\X] \Lambda_2^2 \frac{\partial \X^{0 \mu}}{\partial \dot{A}^b} + \partial^{\mu}\phi_b + \eta^{0\mu}\dot{\phi}_b\right)\X^{0}_{\phantom{0}\mu} \nonumber \\
                             &+ 24\left([\X]\X^{00}-f^{00}\right)V_b \nn \\
                            =& 24 \left\{ (\X^{0}_{\phantom{0}0} + \X^{00})([\X]V_b+\dot{\phi}_b) + (\X^{0\mu}\X^{0}_{\phantom{0}\mu}-f^{00})V_b + \left( \Lambda_2^2 \frac{\partial \X^{0 \mu}}{\partial \dot{A}^b} - \dot{\phi}_b \right)[\X] \right\}
														= 0\,.
\label{eq:HV4}
\end{align}

We can therefore conclude that for any linear combination of the \Pros vector Lagrangians,
\ba
\L_{\K}[A] = \Lambda_2^4 \sum_{n=0}^4 \alpha_n(A^2) \L_n[\K[A]]= \Lambda_2^4 \sum_{n=0}^4 \beta_n(A^2) \L_n[\X[A]]\,,
\label{eq:LKX}
\ea
where the relation between the coefficients $\alpha_n$ and $\beta_n$ is given in \cite{deRham:2014zqa}
the resulting Hessian is of the form
\ba
\mathcal{H}_{ab}= \sum_{n=0}^4 \beta_n \mathcal{H}_{ab}^{(n)}\,.
\ea
Since all the individual Hessians $\mathcal{H}_{ab}^{(n)}$ have the same null direction, with null eigenvector $V^a$, it automatically follows that $V^a$ is also a null eigenvector of the full Hessian $\mathcal{H}_{ab}$ and the full \Pros theory carries a constraint. Remarkably, it is clear from this construction that \Pros theories lie on a different branch of theories as compared to GP theories in terms of how the constraint comes to be implemented. Even though both GP and \Pros are ghost-free theories that carry a constraint, linear combinations of both theories typically break the constraint.

In two dimensions, we can check that the null eigenvector reproduces the exact analytic result  \eqref{eq:L2d}. Recalling  that
\begin{equation}
	\phi^a = x^a + \frac{1}{\Lambda_2^2} A^a\,,
\label{eq:phitoA}
\end{equation}
which then gives
\begin{equation}
V^a=\Z^{0\mu}\p_\mu \phi^a=
	\left(\begin{array}{c}
\Z^{00}(1 - \frac{1}{\Lambda_2^2} \dot{A}_0) - \frac{1}{\Lambda_2^2} \Z^{01} A_0' \\
\frac{1}{\Lambda_2^2} \Z^{00} \dot{A}_1 + \Z^{01}(1 + \frac{1}{\Lambda_2^2} A_1')
\end{array}
	\right)\,.
\label{eq:V0V1}
\end{equation}
Rearranging these terms gives the exact non-perturbative prediction for the two--dimensional eigenvector $v^a$ introduced in  \eqref{eq:eigvec} (up to an irrelevant normalization factor),
\begin{equation}
 V_a = \frac{1}{[\X]}
		\begin{pmatrix}
			2&+&\frac{A_1' - \dot{A}_0}{\Lambda_2^2} \\
			 &-&\frac{\dot{A}_1 - A_0'}{\Lambda_2^2}
		\end{pmatrix}
=  \frac{2}{[\X]}
		\begin{pmatrix}
			1 &+&x/2\\
			&-&y/2
		\end{pmatrix}= \frac{2}{[\X]}v_a
\,.
\label{eq:eigvecbis}
\end{equation}

\section{Kinematics}
\label{sec:appKin}

To perform the scattering amplitudes computations for a given set of polarizations, we need a basis for the polarization vectors $\epsilon_{\mu}^{\lambda}(k_i)$. The polarizations are labelled by $\lambda=-1,0,+1$.

First of all, we consider the center of mass frame where $k_1$ and $k_2$ are traveling in the $\hat{z}$ direction and $k_3$ forms an angle $\theta$ with the $\hat{z}$-axis. We denote the energy by $\omega$ and the norm of the 3-momentum by $k$
\begin{align}
	k_1^{\mu} &= (\omega, 0, 0, k) \label{eq:k1 }\\
	k_2^{\mu} &= (\omega, 0, 0, -k) \label{eq:k2} \\
	k_3^{\mu} &= (\omega, k \sin(\theta), 0, k \cos(\theta)) \label{eq:k3} \\
	k_4^{\mu} &= (\omega, - k \sin(\theta), 0, - k \cos(\theta))\,. \label{eq:k4}
\end{align}
In this set-up the polarization vectors basis can be chosen to be
\begin{center}
\begin{tabular}{ c c c }
	$\epsilon_{\mu}^{+}(k_1) = \begin{pmatrix} 0 \\ 1 \\ 0 \\ 0 \end{pmatrix}$ & $\epsilon_{\mu}^{-}(k_1) = \begin{pmatrix} 0 \\ 0 \\ 1 \\ 0 \end{pmatrix}$ & $\epsilon_{\mu}^0(k_1) = \begin{pmatrix} -\frac{k}{m} \\ 0 \\ 0 \\ \frac{\omega}{m} \end{pmatrix}$ \\
	$\epsilon_{\mu}^{+}(k_2) = \begin{pmatrix} 0 \\ -1 \\ 0 \\ 0 \end{pmatrix}$ & $\epsilon_{\mu}^{-}(k_2) = \begin{pmatrix} 0 \\ 0 \\ 1 \\ 0 \end{pmatrix}$ & $\epsilon_{\mu}^0(k_2) = \begin{pmatrix} -\frac{k}{m} \\ 0 \\ 0 \\ -\frac{\omega}{m} \end{pmatrix}$ \\
	$\epsilon_{\mu}^{+}(k_3) = \begin{pmatrix} 0 \\ \cos(\theta) \\ 0 \\ -\sin(\theta) \end{pmatrix}$ & $\epsilon_{\mu}^{-}(k_3) = \begin{pmatrix} 0 \\ 0 \\ 1 \\ 0 \end{pmatrix}$ & $\epsilon_{\mu}^0(k_3) = \begin{pmatrix} -\frac{k}{m} \\ \frac{\omega}{m} \sin(\theta) \\ 0 \\ \frac{\omega}{m} \cos(\theta) \end{pmatrix}$ \\
	$\epsilon_{\mu}^{+}(k_4) = \begin{pmatrix} 0 \\ -\cos(\theta) \\ 0 \\ \sin(\theta) \end{pmatrix}$ & $\epsilon_{\mu}^{-}(k_4) = \begin{pmatrix} 0 \\ 0 \\ 1 \\ 0 \end{pmatrix}$ & $\epsilon_{\mu}^0(k_4) = \begin{pmatrix} -\frac{k}{m} \\ -\frac{\omega}{m} \sin(\theta) \\ 0 \\ -\frac{\omega}{m} \cos(\theta) \end{pmatrix}$\,.
\end{tabular}
\end{center}
One can verify that this basis satisfies the polarization vector properties, for a given vector $k_i$ (\ie $i=1,...,4$ fixed)
\begin{align}
	\epsilon_{\mu}^{\lambda}(k_i) k_i^{\mu} &= 0 \label{eq:keps0} \\
	\epsilon_{\mu}^{\lambda}(k_i) \epsilon^{\mu,\lambda'}(k_i) &= \delta^{\lambda \lambda'} \label{eq:epseps0} \\
	\sum_{\lambda=-1}^1 \epsilon_{\mu}^{\lambda}(k_i) \epsilon_{\nu}^{\lambda}(k_i) &= \eta\mn + \frac{{k_i}_{\mu} {k_i}_{\nu}}{m^2}\,. \label{eq:completeness}
\end{align}
We also have the following kinematical constraints
\begin{align}
	\omega &= \frac{\sqrt{s}}{2} \label{eq:omegatos} \\
	k &= \frac{1}{2}\sqrt{s-4m^2} \label{eq:ktos} \\
	t &= -\frac{1}{2}(s-4m^2)(1-\cos(\theta)) \label{eq:ttos} \\
	u &= -(s+t)+4m^2 \,, \label{eq:utos}
\end{align}
which enable us to fully specify the kinematics with the two parameters $(s,\theta)$.

\section{Resummation of the complete DL of massive gravity}
\label{sec:appDL}

In this Appendix we provide an explicit formula resumming the DL of massive gravity to all orders in $\Phi\mn = \p_{\mu} \p_{\nu} \phi$.
For convenience, we work here in the formulation of the theory in terms of the tensor $\X$ as in \eqref{eq:LKX} and we only need to focus on the contribution of the $\beta_n$ which is independent of $A$, so in what follows we may consider the $\beta_n$'s to be constant.
As derived by Ondo and Tolley in \cite{Ondo:2013wka}, the scalar--vector sector of this DL is
\begin{align}
\label{eq:DLAndrew}
	\L_{\text{DL}} = \left\{ \vphantom{\frac{1}{2}} \right. &- \frac{\beta_1}{4} \left( \frac{1}{2} F^a_{\mu} \omega^b_{\phantom{b}\nu} \delta^c_{\rho} \delta^d_{\sigma} + (\delta + \Phi)^a_{\mu} \delta^b_{\nu} \left[ \omega^c_{\phantom{c}\rho} \omega^d_{\phantom{d}\sigma} + \frac{1}{2} \delta^c_{\rho} \omega^d_{\phantom{d}\alpha} \omega^{\alpha}_{\phantom{\alpha}\sigma} \right] \right) \\
											                                &- \frac{\beta_2}{8} \left( 2 F^a_{\mu} \omega^b_{\phantom{b}\nu} (\delta+\Phi)^c_{\rho} \delta^d_{\sigma} + (\delta + \Phi)^a_{\mu} (\delta + \Phi)^b_{\nu} \left[ \omega^c_{\phantom{c}\rho} \omega^d_{\phantom{d}\sigma} + \delta^c_{\rho} \omega^d_{\phantom{d}\alpha} \omega^{\alpha}_{\phantom{\alpha}\sigma} \right] \right) \nonumber \\
														                                &- \left. \frac{\beta_3}{24} \left( 3 F^a_{\mu} \omega^b_{\phantom{b}\nu} (\delta+\Phi)^c_{\rho} (\delta+\Phi)^d_{\sigma} + (\delta + \Phi)^a_{\mu} (\delta + \Phi)^b_{\nu} (\delta + \Phi)^c_{\rho} \omega^d_{\phantom{d}\alpha} \omega^{\alpha}_{\phantom{\alpha}\sigma} \right) \right\} \epsilon^{\mu \nu \rho \sigma} \epsilon_{abcd}\,,\nn
\end{align}
where $\omega$ is a composite field defined by
\begin{equation}
	\omega = \sum_{n,m} \frac{(n+m)!}{2^{1+n+m}n!m!}(-1)^{n+m} \Phi^n F \Phi^m\,.
\label{eq:omega}
\end{equation}
The expression \eqref{eq:DLAndrew} has the advantage to be compact and complete but it is useful to rewrite it only in terms of $F$ and $\Phi$, the actual field content of the theory. It can be proven by basic binomial manipulations that the complete DL of massive gravity can be resummed to all orders in the following way
\begin{align}												\label{eq:resumfullDL}
	\L_{\text{DL}} =& - \frac{\beta_1+2\beta_2+\beta_3}{8}F\mnup F\mn + \frac{\beta_1+4\beta_2+3\beta_3}{8\Lambda_3^3}F^{\mu \alpha}F^{\nu}_{\phantom{\nu} \alpha} \Phi\mn - \frac{\beta_2+\beta_3}{8\Lambda_3^3}F\mnup F\mn[\Phi] \\
										 & + \sum_{p=2}^{\infty} \sum_{k=0}^{p} \frac{1}{\Lambda_3^{3p}} \left\{ \frac{\beta_1+4\beta_2}{8} - \frac{\beta_3}{8} \left( 8p - 23 - 4 \frac{k(p-k)(4p-9)}{p(p-1)} \right) \right\} \frac{(-1)^p}{2^p} {{p}\choose{k}} [F \Phi^{k} F \Phi^{p-k}] \nonumber \\
										 & + \sum_{p=2}^{\infty} \sum_{k=0}^{p-1} \frac{1}{\Lambda_3^{3p}} \left\{ - \frac{\beta_2}{4} + \frac{\beta_3}{4} \left( 2p - 9 + 2 \frac{(p-k-1)^2+k^2}{p-1} \right) \right\} \frac{(-1)^p}{2^p} {{p-1}\choose{k}} [F \Phi^{k} F \Phi^{p-k-1}]S_1(\Phi)
												\nonumber \\
										 & + \sum_{p=2}^{\infty} \sum_{k=0}^{p-2} \frac{1}{\Lambda_3^{3p}} \left\{ - \frac{\beta_3}{4} \left( 2p - 5 \right) \right\} \frac{(-1)^p}{2^p} {{p-2}\choose{k}} [F \Phi^{k} F \Phi^{p-k-2}]S_2(\Phi)
												\nonumber \\
										 & + \sum_{p=3}^{\infty} \sum_{k=0}^{p-3} \frac{1}{\Lambda_3^{3p}} \left\{ \frac{\beta_3}{6} \left( p - 2 \right) \right\} \frac{(-1)^p}{2^p} {{p-3}\choose{k}} [F \Phi^{k} F \Phi^{p-k-3}]S_3(\Phi)\,, \nonumber
\end{align}
where the $S_n(\Phi)$ are a short-hand notation for
\begin{align}
	S_1(\Phi) &= [\Phi] \\
	S_2(\Phi) &= [\Phi]^2 - [\Phi^2] \\
	S_3(\Phi) &= [\Phi]^3 - 3[\Phi][\Phi^2] + 2[\Phi^3]\,.
\label{eq:SnPhi}
\end{align}
Here we use brackets as a notation for the trace. Some terms of the expansion \eqref{eq:resumfullDL} might include contributions of the form $[F^2]$, which really stands for the trace of the square of the field-strength tensor $F\mn$. In this case, the convention is opposite to the one introduced in \eqref{eq:LK3}. Indeed,
\begin{equation}
	[F^2] = F\mnup F_{\nu \mu} = - F\mnup F\mn
\label{eq:newconv}
\end{equation}
Note that the coefficients $\beta_n$  are not linearly independent, indeed they satisfy
\begin{equation}
	\beta_1 + 2 \beta_2 + \beta_3 = 2\,.
\label{eq:betas}
\end{equation}
Expanding \eqref{eq:resumfullDL} up to quartic order and using \eqref{eq:betas} to eliminate $\beta_3$ gives
\begin{align}
	\L_{{\text{MG DL}}}^{(2)} =& -\frac{1}{4}F\mnup F\mn
		\label{eq:LMGDL2} \\
	\L_{{\text{MG DL}}}^{(3)} =& - \frac{2 - \beta_1 - \beta_2}{8} F\mnup F\mn \Box \phi  + \frac{3 - \beta_1 - \beta_2}{4} F^{\mu \alpha}F^{\nu}_{\phantom{\nu}\alpha} \p_{\mu} \p_{\nu} \phi
		\label{eq:LMGDL3} \\
	\L_{{\text{MG DL}}}^{(4)} =& - \frac{2 - \beta_1 - 2 \beta_2}{16} F\mn^2\left( (\Box \phi)^2 - (\p_{\alpha} \p_{\beta}\phi)^2 \right) - \frac{7 - 3 \beta_1 - 5 \beta_2}{8}F^{\mu \alpha} F^{\nu}_{\phantom{\nu} \alpha} \p_{\mu} \p^{\beta} \phi \p_{\nu} \p_{\beta} \phi \nonumber \\
								             & + \frac{6 - 3 \beta_1 - 5 \beta_2}{8} F^{\mu \alpha} F^{\nu}_{\phantom{\nu} \alpha} \p_{\mu} \p_{\nu} \phi \Box \phi - \frac{5 - 2 \beta_1 - 3 \beta_2}{8}F\mnup F\abup \p_{\mu} \p_{\alpha} \phi \p_{\nu} \p_{\beta} \phi
			\label{eq:LMGDL4}
\end{align}
Comparing \eqref{eq:LKDL2}--\eqref{eq:LKDL4} to \eqref{eq:LMGDL2}--\eqref{eq:LMGDL4}, and using the relation between the coefficients $\alpha_n$ and $\beta_n$ as provided in \cite{deRham:2014zqa} one can see that the vector-scalar sector of our new Proca interactions in the DL exactly coincides with this sector in the DL of massive gravity.
%

\bibliographystyle{JHEP}
\bibliography{references}

\end{document}